\documentclass{article}
\usepackage{frascatiphys,here,graphicx,subfigure}
\begin{document}
\title{ METHODS AND MODELS FOR HADRON PHYSICS \\ Round Table}
\author{%
\hspace{-5mm}
\begin{minipage}{120mm}
\begin{tabular}{rl}
Panelists:&\hspace{-4mm} C. Davies, S.   Faccini, H. Lipkin,
L. Maiani (Chair),\\&\hspace{-4mm}  F. J. Yndur\'ain\\  
Contributors:&\hspace{-4mm} C. Bugg, S. Eidelman, P. Faccioli, 
S. Glazek, Y. Glozman,\\&\hspace{-4mm} E. Klempt, H. Koch, 
J. Lee-Franzini, R. Mussa, E. Pallante,\\&\hspace{-4mm}   
S. Paul, K. Seth, U. Wiedner\\
Convenors:&\hspace{-4mm} M. P. Lombardo, S. Miscetti, S. Pacetti\\
\end{tabular}
\end{minipage}
}
\maketitle
\baselineskip=11.6pt
\begin{abstract}
A round table held during the Hadron07 Conference focusing
on experimental observations of new hadronic states, 
on theoretical perspectives for their description, and on 
the role of hadronic spectroscopy in furthering our knowledge 
of the fundamental theory of strong interactions. 
\end{abstract}
\baselineskip=14pt
\section{Opening Statements}

\vskip .5 truecm
\noindent
\underline{L. Maiani}

As I have already given the introductory review this morning,
I will just invite Professor Yndur\'ain to begin, 
and Professor Davies and Doctor Faccini to continue after him.
Which are the problems that you would like to put to the attention
of the audience?

\vskip .5 truecm
\noindent
\underline{F. J. Yndur\'ain}

Since, for obvious reasons of age, I imagined I was going to be the first panelist to talk, I have prepared a list of questions, experimental and theoretical, that I would like to have solved or, 
at least, understand them better.

\noindent 1)\quad  We are all convinced that the particle $\eta_b$ exists, but I for one 
will have nagging doubts until it is actually discovered. Particularly since there are sound 
theoretical calculations of its mass (some 35 MeV lighter than the upsilon), so one could check 
ideas in QCD for bound states. I am aware that this is not an easy experiment, but there you are.

\noindent 2)\quad One of the mysteries of QCD is the extent to which the {\sl  constituent}
quark model works. I mean, it is OK to say that as quarks move in the soup of 
gluons and quark-antiquark pairs in a hadron they acquire an effective mass of some 
300~MeV; but, except for the Goldstone mesons, this simple model works much better than what it should. For example, the relations of 
total cross sections 
\[\sigma_{\pi\pi}:\sigma_{\pi N}: \sigma_{NN}=2\times 2:2\times 3: 3\times3,
\]
$\sigma_{\pi N}=\sigma_{KN}$, etc. work at the level of 10\%. Yet they are obtained 
assuming that hadrons consist of only constituent quarks, that behave as if they were free.
These relations were obtained in the sixties of last century, and we are nowhere near understanding them; for example, they are contrary to what one finds in deep inelastic scattering, 
where hadron structure functions have a strong gluonic component.

\noindent 3)\quad We have a challenge in obtaining the pion-pion scattering amplitudes at low energy. 
Much improvement has been achieved recently, particularly for the S0 wave thanks to precise measurements of 
two-pion and three-pion  kaon decays here at Frascati, and of $K_{e4}$ 
decays by the NA48/2 collaboration. 
In this way, one can start to test predictions of chiral perturbation theory, and contribute to the 
construction of  very precise $\pi\pi$ scattering amplitudes.

\noindent 4)\quad Of course, the resonances found in charmonium (the $X$, $Y$, $Z_s$) 
have shown a rich structure that ought to be investigated further.

\noindent 5)\quad (This in response to a question from the audience). I would like to remark that 
a much-publicised ``discrepancy" between the pion form factor as measured in $e^+e^-\to\pi\pi$ 
and in $\tau^-\to\nu\pi^-\pi^0$ is not incomprehensible nor does it 
pose a problem for incorporating  $\tau^-\to\nu\pi^-\pi^0$ results into e.g., calculations 
of the muon $g-2$. All one has to do is to take into account that the rho states contributing 
there are different, $\rho^0$ in the first case and $\rho^-$ in the second. And, because 
the rho contribution is so large (about a factor 50 at the peak) even a small mass and 
width  difference between the two produces a large difference in the form factors. 
In fact, one can make the calculation and, once this effect is taken into account, 
the discrepancy between the  pion form factor  in $e^+e^-\to\pi\pi$ 
and in $\tau^-\to\nu\pi^-\pi^0$ is quite compatible with the systematic normalization 
uncertainties in these processes [see e.g. F.~Jegerlehner, Proc. Int. Frascati Conf., 2003,  hep-ph/0310234; 
S.~Ghozzi and F.~Jegerlehner, Phys. Lett. B {\bf 583}, 222 (2004);  
J. F. de Troc\'oniz, and F. J., Yndur\'ain, Phys. Rev. D {\bf 71}, 073008 (2005)].

\vskip .5 truecm
\noindent
\underline{H. Lipkin} \footnote{H. Lipkin could not participate and
sent  his contribution via e-mail.}

 We still have a great deal to learn about how QCD makes hadrons out of
 quarks and gluons. We don't know enough about QCD to believe any hadron
 model.  All the theoretical approaches including the lattice have drastic
 oversimplifications which leave us still far from our goal.

      The following questions may lead to a better understanding of how
 hadrons are made from quarks and gluons.

 1) What is the constituent quark picture? There are several versions.

 2) Where does a particular version work very beautifully? Where does it
  not work so beautifully? Where does it not work at all?
 
3) Why?

 Most theoretical treatments start with well defined models with a number
 of free parameters and try to use the data to fix the parameters. We look
 for clues in the data, for puzzles that challenge the conventional wisdom.

 Our approach is very different from that of few-body nuclear physics which
 begins with a system of particles whose masses and interactions are
 assumed to be known. Our version of the constituent quark model begins
 with constituent quarks whose nature, masses and structures are not known,
 have an unknown dynamical origin, may differ between different hadrons and
 have so far not been explained by QCD. One challenge we face is how to use
 the new data on heavy quark hadrons to find clues to the nature, masses
 and structures of these constituent quarks.

 Our quark masses are effective masses which contain contributions from
 complicated interactions in ways that are not understood. The fact that
 the same values for these effective quark masses are found in experimental
 masses of mesons and baryons is a striking challenge to all attempts to
 construct a more basic microscopic theory. This may indicate a new yet
 undiscovered symmetry or supersymmetry. We go far beyond conventional
 quark model investigations which use either a nonrelativistic or
 relativistic few-body model with fixed mass parameters.  Lattice QCD has
 been so far unable to get the kinds of predictions between mesons and
 baryons that have been obtained with this phenomenological constituent
 quark model.

 We start with experimental facts and surprising agreements from models
 with very simple assumptions. We want to find how maximum agreement with
 experiment can come from minimum assumptions.

\noindent
{\em A.  Hadron Mass predictions and relations.}

     1) The simplest assumptions were first proposed by Sakharov and
 Zeldovich and later independently discovered by me. The hadron mass
 operator consists of (1) an effective quark mass containing all the
 spin-independent contributions including potential and kinetic energies.
 and (2) a two-body hyperfine interaction proportional to $\sigma_i \sigma_j$.
 
    a) The difference between the effective mass contributions for any
 two flavors is the same for all ground state hadrons, both mesons and
 baryons.

     b) The ratio between the hyperfine energies for any two combinations
 of flavors is the same for all ground state hadrons, mesons and baryons.
     The number of experimental regularities that follow from these
 simple assumptions is striking, and leads to the remarkable results in our
 paper hep-ph/0611306.

     We compare a meson consisting of a valence quark of flavor i and a
 light quark system having the quantum numbers of a light antiquark with a
 baryon consisting of a valence quark of the same flavor i and a light
 quark system having the quantum numbers of a $ud$ diquark of spin $S$.
     We make no assumption about the nature and structure of the valence
 quark or the light quark system. Our results apply not only to simple
 constituent quark models but also to parton models in which hadrons
 consist of valence current quarks and a sea of gluons and quark-antiquark
 pairs.
     When the hyperfine interaction between the quark and the antiquark or
 diquark is taken out (a simple procedure with no free parameters), the
 baryon-meson mass difference is independent of the flavor i of the quark
 which can be $u$, $s$, $c$ or $b$. This alone is a striking challenge for QCD
 treatments which so far have not found anything like this regularity
 between meson and baryon masses.  These results can be seen in
 hep-ph/0611306.

     2) The next version is that of DeRujula, Georgi and Glashow which
 assumes that the two-body hyperfine interaction is inversely proportional
 to the product of the two effective quark masses. The magnetic moments of
 the quarks are inversely proportional to same effective quark masses. This
 gives the remarkably successful predictions for the magnetic moments of
 the neutron, proton and Lambda and also some new mass relations.

\noindent
{\em      B. Use of constituent quarks in weak decays.}

 Most treatment of weak decays assume that the weak transition is only
 between the valence quarks of the initial and final states. Quark diagrams
 are classified and assumptions are made like factorisation, etc. which
 neglect certain diagrams.

 Here again I look for simple relations that work based on simple
 approximations like the following:

   1) Charmless strange B decays are assumed to be dominated by the
 penguin diagram. The discovery of CP violation in these decays indicates
 that there must be some other amplitude that interferes with the penguin.
 In the limit where there is only a penguin contribution the four
 independent branching ratios for $B\to K\pi$ decays are all related and
 proportional to the penguin amplitude.  We define three linear
 combinations of the four branching ratios which vanish if there is only
 the penguin contribution. Any linear combination differing from zero
 provides a clue to an additional contribution which interferes with the
 penguin.
    Before recent new more precise experimental data were available all
 three of these linear combinations were statistically consistent with
 zero. But new data show two of the three to be appreciably different,
 while one of them is still consistent with zero. If this is correct it
 tells us something about which contributions are producing the CP
 violation. It may imply a cancellation that makes one of these
 contributions vanish. We now need more and better data to check this out.
 It is on the Los Alamos Archive at hep-ph/0608284.

    In vector-pseudoscalar charmless strange $B$ decays like $K-\rho$ a
 phenomenological parity selection rule agrees surprisingly well with the
 data, but does not agree with simple models. But it comes from a simple
 description using hadron spectroscopy. The dominant penguin diagram
 produces a strange antiquark, a u or d spectator quark, and gluons. In
 flavor SU(3) this state must be in an octet with the isospin and
 strangeness quantum numbers of a kaon. There are only two possible states
 with these quantum numbers, a normal kaon which is a quark-antiquark pair
 and an "exotic kaon" which has the opposite generalised charge conjugation
 and cannot be constructed from a single quark-antiquark pair. The data are
 consistent with a model which excludes the exotic kaon. This fits
 naturally into a picture where the strong interaction scattering producing
 the final state is dominated by intermediate resonances and these all have
 normal quantum numbers so the exotic contribution is suppressed. This is
 discussed in my paper hep-ph/0703191. But this exotic suppression ansatz
 does not make sense in the standard treatments which only assume states of
 two quark-antiquark pairs and no multiparticle intermediate states.

\vskip .5 truecm
\noindent
\underline{C. Davies}
%
%

Lattice QCD is now able to calculate precise values for gold-plated 
hadron masses. These are summarised in the "ratio plot" of lattice 
QCD/experiment in my talk. Gold-plated means stable, well away 
from decay thresholds and accurately measured experimentally. 
For the calculations in lattice QCD that have been done so far 
there is excellent agreement with experiment when realistic sea 
quark effects are included in the calculation. This is after having 
fixed the 5 parameters of QCD for these calculations 
(4 quark masses and a coupling constant) 
from 5 other hadron masses. For example, $D$ and 
$D_s$ masses have been calculated, and agree with experiment, to 7 MeV. 
This level of accuracy would be impossible in any approximate model of QCD and 
is a very stringent test of the theory.  So lattice QCD is now 
testing QCD.
Most of the calculations have been done for mesons so far since they 
are easier. Gold-plated baryon masses will be calculated in 
the next few years and these will provide additional tests of QCD. 

The masses of excited and unstable states are not nearly so easy 
to calculate. The precision possible from a lattice QCD calculation 
will not be as good. There are still interesting results to be had 
from doing the calculations, but you need to decide what question 
is being asked i.e. what level of precision is needed to answer it? 
You also then need to pay attention to the sources of systematic 
error in lattice calculations of these states. 

One interesting lattice calculation underway is that of the baryon spectrum 
by the LHPC collaboration. The ground-state nucleon is gold-plated - the 
other states are not, and some are very broad and poorly known 
experimentally. A basic issue here is exactly how many states there are, 
and it is one that experimentalists are tackling. 
The LHPC collaboration is beginning preliminary tests on quenched 
gluon field configurations (i.e. not including the effects of 
sea quarks) of the kinds of operators, lattice volumes etc that they 
would need for a complete lattice calculation. 
They have obtained approximate masses for a lot of states, so it is encouraging 
news that this calculation is possible. The quenched results may be accurate enough
(with, say, 20\% systematic errors) to answer some of the interesting issues. 
On including sea quark effects, multihadron states in the spectrum 
will be an additional problem and it is not clear how well that can be 
tackled. It may obscure some of the masses you would like to extract 
even further and will certainly make quantitative analysis as a function of
light quark mass very hard. 
\begin{figure}[h!]
\centering\vspace{-8mm}
\includegraphics[%
  scale=0.48,
  angle=0,
  origin=c]{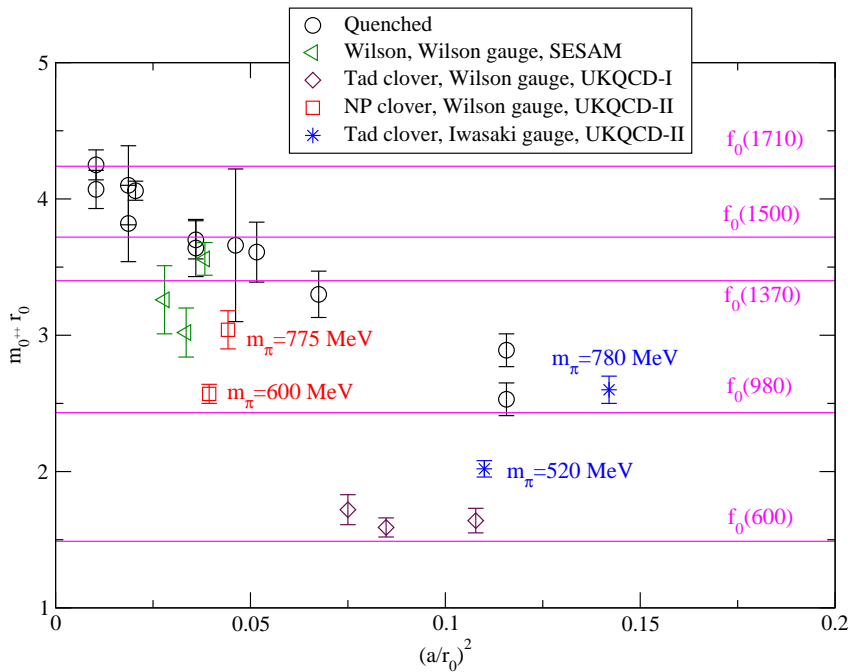}
\vspace{-1cm}
\caption{\it Summary of unquenched results for lightest flavour
singlet $0^{++}$ mesons, from McNeile, Lat07.
The unquenched results are from SESAM, and UKQCD.}
\vspace{-5mm}
\end{figure}\\
Flavor singlet/glueball masses are even harder - see the plenary talk by 
C.~McNeile at LAT07, 0710.0985[hep-lat]. A summary of the lattice results 
contrasted with some experimental meson masses is reproduced below. 
The key to calculations in this area will be very high statistics, i.e. 
fast sea quark formalisms, and a good operator basis, so that all the mixing 
issues can be handled. \\
{\bf Summary} 

1) High precision results for gold-plated hadrons will continue to improve. 
These are the ones that provide the stringent tests of QCD because of the 
accuracy that is possible. For example, accurate simultaneous (i.e. 
with only one set of quark masses and coupling constant) calculations 
of heavy-heavy, heavy-light and light-light meson masses are now possible
in lattice QCD and could not be done in any derived model of QCD. 

2) This needs to be extended to gold-plated baryons and to 'silver-plated' 
mesons (particles that are unstable but relatively narrow like the phi, D* etc). 
Eventually there will also be results for higher-lying and more unstable 
particles. The same level of accuracy will not be possible, however.

3) The same remarks apply to electroweak decay rates. The gold-plated ones 
are those having at most one gold-plated hadron in the final state. We now have $f_D$ and 
$f_{D_s}$ to 2\%. Calculations are in progress for $\Gamma_{e^+e^-}$ for $J/\psi$ 
and $\phi$. This is having a strong impact on the flavor physics programme. 
We also expect accurate form factors for semileptonic decay 
and structure function moments for 
baryons to be possible. 

4) It is important to test lattice QCD with different quark formalisms 
and more results from a variety of formalisms will become available over 
the next few years.


\vskip .5 truecm
\noindent
\underline{R. Faccini}

There are several areas where flavour physics can probe strong interactions
and therefore verify or falsify models and lattice calculations:
\begin{itemize}
\item  fits to the unitarity triangle parameters, $\bar{\rho}$ and
$\bar{\eta}$. The current accuracy of the experimental measurements is such
that the implications of the measurement of $\epsilon^{\prime}$ in kaon
decays is entirely dominated by the theoretical uncertainties and that  the
other quantities measured on the lattice can be overconstrained by other
experimental measurements, if the Standard Model is assumed. This implies,
as detailed in the dedicated publication~[M.~Bona {\it et al.}  [UTfit Collaboration],
  JHEP {\bf 0610} (2006) 081], that before
bringing a significant contribution to the unitarity triangle the
measurements on lattice of $f_{B_d}$ and $f_{B_s}$ must improve by at least
a factor three.
On the other side the current measurements are still critical for the
interpretations of the current data that include physics beyond the Standard
Model.
\item Semileptonic $B$, $D$ and $K$ decays. The best probes of QCD come in
these systems where only two quarks interact. Inclusive measurements are
particularly dependent on the availability of models that describe the data
and can also be used to probe the parton-hadron duality assumed in all
predictions. Exclusive measurements rely instead on the availability of the
form factors. The statistics is high enough to allow the data to constraint
the $q^2$ dependencies, and can therefore often discriminate among
theoretical models that  estimate the overall normalization.
\item Most of the techniques to measure weak phases exploit the interference
between amplitudes that have both weak and strong phases. The best
environment to apply such techniques are the multibody decays, and their
actual success depends on the possibility of properly modelling the strong
phases of multibody decays. Several approaches have been developed in the
past decades to take this problem (isobar model, K-Matrix,\ldots) but there
is still large arbitrariness in this kind of analyses. The field would
profit from a systematic study that gives precise rules on the approach to
follow and the resonances/poles to consider. There is an increasing wealth
of experimental results of direct production of light mesons~[see the
contribution from C. Bini at this Conference]
that must be used to support such a study.
\item Heavy quarkonium spectroscopy. The spectroscopy of the bound states of
a pair of heavy quarks can be predicted with relatively good accuracy with
potential models~[N. Brambilla {\it et al.} [Quarkonium working Group],
hep-ph/0412158] This makes this field a good ground to observe
new forms of aggregation. Indeed there have been recently a large number of
experimental evidences that QCD does not only bind quark pairs but also
groups of four quarks or of two quarks and gluons~[Contribution from R. Mussa 
at Hadron'07 and from R. Faccini at Lepton Photon '07 (arXiv~0801.2679)]. 


The path towards the full understanding of the new spectroscopy is still
long, both from the theoretical and the experimental point of view. In
particular as far as the latter is concerned, only a very small fraction of
possible final states and production mechanisms have been studied on the
data available from B-Factories. Finally some of the measurements, in
particular those implying $D$-meson reconstruction, will require a
significantly larger statistics than what the current generation of
experiments will collect. 

The diagram below summarizes the current status

 \end{itemize}

\begin{figure}[htb]
\includegraphics[width=13 truecm, angle=0]{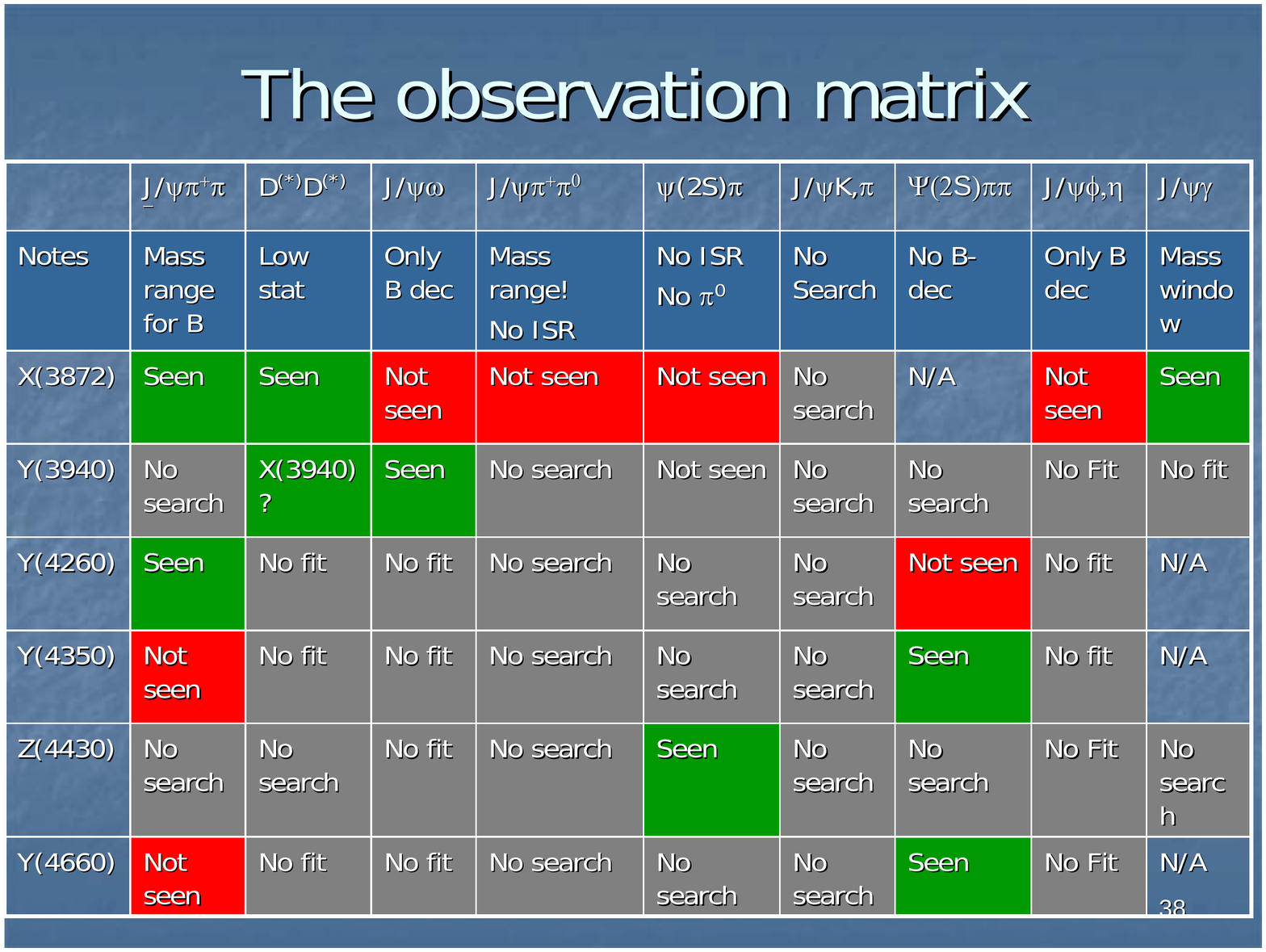}
\end{figure}

\vskip .5 truecm
\noindent
\section{Discussion}

\vskip .5 truecm
\noindent
\underline{L. Maiani}
We can now open the general discussion. It might be useful perhaps
to divide the rest of the session in two, one devoted to experiments
and the other to theory.

\vskip .5 truecm
\noindent
\underline{K. Seth}
I am an experimentalist and I have a question for the theorists. When
Lattice came out we thought that experimentalists will soon be out of
business. Now I feel our existence is not so much threatened, because now
when I ask a question they occasionally reply that they cannot handle it.
I give you an example. When I ask if they can calculate the timelike form
factors of hadrons (protons, pions, kaons) which we measure so
painstakingly, they tell me that they cannot handle quarks in Euclidean
time. Is this really true that there are things which cannot be measured
on the Lattice?

\vskip .5 truecm
\noindent
\underline{C. Davies}
Yes. Lattice QCD calculations serve a number of useful purposes but 
they were never intended to put experimentalists out of business, 
more as away of providing good tests of QCD against experiment in 
hadron physics. It is true that there are also only certain things 
that we can calculate.

\vskip .5 truecm
\noindent
\underline{K. Seth}
Thank you for being honest! (smiles).

\vskip .5 truecm
\noindent
\underline{L. Maiani}
I have  a comment: I think that the purpose of Lattice Gauge Theory
is to make explicit what QCD can say, but certainly this does not imply
that experimentalists have to go away. What we would like to do
is validate QCD by comparing the results of lattice calculations with
experiments. Christine Davies was very honest, but even if things such as
timelike form factors and multiquark resonances were measured on the
lattice,
this certainly should not prevent the experimentalists from challenging
such predictions. Consider charmonium: this is a case in which potential
models in principle should work. We can compute everything, and yet we find
different things by doing experiments. This is what is meant for discovery.

\vskip .5 truecm
\noindent
\underline{K. Seth}
There are other cases where I would very much like to have help from
Lattice. For example, two-photon decays of charmonia. In the leading order
this is a pure QED process. QCD comes in in terms of strong radiative
corrections. Unfortunately they are only available at the one loop level,
with corrections as large as 100\%. Obviously, these cannot be trusted. Can
Lattice help?

\vskip .5 truecm
\noindent
\underline{C. Davies}
There have been exploratory calculations at 
Jefferson Lab on charmonium two-photon decays and it is certainly 
on the list of things that we can do better. Potential 
models can do well for bottomonium and charmonium 
but when we look in any detail we find discrepancies with 
experiment for a given potential, especially if
we are not allowed to readjust the parameters, and 
if we look at decay rates as well as the spectrum. 
Lattice QCD calculations can provide a big improvement 
in accuracy here, which is important because the 
experimental results are very accurate. And we 
can also predict heavy-light physics with no additional 
parameters (because we are doing real QCD). 

\vskip 0.5 truecm
\noindent
\underline{L. Glozman}
First I  would like to make a comment on a statement by our chairman and
by prof . Yndur\'ain that quark models work extremely well. It might work
well
for spectroscopy of the ground state, but we know that it fails for
excited states. 
My second point concerns mass generation, confinement and chiral symmetry 
breaking. Heavy quarks do not allow the study of chiral symmetry. 
To address these issues it is important to consider light quark
spectroscopy, which is not well represented in the plenary program
at this Conference. Moreover, light quarks have probably an impact on the 
understanding of the QCD phase diagram.

\vskip 0.5 truecm
\noindent
\underline{U. Wiedner}
I would like to correct a statement that hadron machines are not
very useful for spectroscopy. 
For instance  Antiproton proton machines yield the 
most precise charmonium states.

\vskip 0.5 truecm
\noindent
\underline{R. Faccini}
I haven't said they did not contribute, but, simply, that they importance
is limited.

\vskip 0.5 truecm
\noindent
\underline{U. Wiedner}
Panda experiment at Fair will address this problem 
Theorists should tell us the precise tetraquark spectrum, would
be good if lattice and models would come together in order
to give some guidance to the experiments.

\vskip 0.5 truecm
\noindent
\underline{J. Lee Franzini}
I think we have lots of light quark presentations, in particular
from DAFNE. This is in response to Glozman comment.
I also have a question to Christine: can lattice calculate $g-2$?

\vskip 0.5 truecm
\noindent
\underline{C. Davies}
There are lattice calculations addressing the 
hadronic contribution, still exploratory, led by Tom 
Blum at BNL. The time scale is hard to estimate since 
it is a difficult calculation.

\vskip 0.5 truecm
\noindent
\underline{F. J. Yndur\'ain}
Lattice results for the $g-2$ are not for tomorrow, and it will also 
be difficult to improve on the naive models for light by light scattering.

\vskip 0.5 truecm
\noindent
\underline{L. Maiani}
I would like to come back to Juliet's concern. This problem is
going to be solved in a different way. 
You want to know $g-2$ to a high precision to understand if there
are deviations from the Standard Model. But if such deviations are there,
the LHC will see the new particles.
So I think LHC will see such effects before we solve the light by light
scattering, which is an interesting problem, but after all
I agree with what Paco said, it is not for tomorrow.

\vskip 0.5 truecm
\noindent
\underline{D. Bugg} 
I would like to
follow up to Glozman's remark. We would like much more data.
Statistics on charmonium order of hundred events after ten years. 
Many of these big detectors are now running to the end of present
experiments. 
Big detectors should go in a pion beam. If you do
so you can collect an enormous statistics. You need a polarized target,
and you can get  roughly one million event per day, and all
good events. You can cover the  entire mass spectrum up to 2.5 GeV in one year.
I would like to see JPARC or other machines with a pion beam do such an 
experiment.
Beam intensity is 'trivial'. All of the baryon and
meson spectroscopy up to third excitations could be 
calculated in a couple of years run.

\vskip 0.5 truecm
\noindent
\underline{P. Faccioli}
In this discussion it has been recognized the importance of
understanding the interrelation between 
Quark model and QCD. Other models have been examined in the 
past few years, and we have learned that dynamical symmetry
breaking produces a dynamical mass.

\vskip 0.5 truecm
\noindent
\underline{E. Klempt}
Chiral symmetry breaking is certainly responsible for mass generation, 
but how about constituent  quark mass for the excited states?
I also agree with Bugg, that we need new data.  But we can also use 
the available data from $B$-factories. It is very important that
people publish the data in such a way that everybody can use them for further
analysis. We (Crystal Barrel) have set up WEB pages from where
you can get lots of information.

\vskip 0.5 truecm
\noindent
\underline{L. Maiani}
I this point I feel I would like to make a comment.
I think the table shown by Faccini [the table was still
displayed on the screen, note of the Editor] is illuminating,
and helps answering  a previous question about guidance from theory.
Guidance from theory is difficult to get, if you do not have data
to start with.  The question is, 
how long it is going to take to fill all the 'grey boxes' in the table,
i.e. all gap in the spectroscopy, and which machines can do that? 
Are the present facilities enough or not?

\vskip 0.5 truecm
\noindent
\underline{K. Seth}
The problem is that so far these new states have been populated mainly by
$B$-decays or double charmonium production at the B factories. Even with
several hundred femtobarns of luminosity in most case the event statistics
is terribly small. They would need several attobarn to improve the
situation in any significant way. In one case, for $X(3872)$ the Tevatron
has shown that if a completely different way of reaching these states can
be found, one can make a real difference. We just have to find other ways
than the $B$-factory ways of populating these exotic states.

\vskip 0.5 truecm
\noindent
\underline{L. Maiani}
Can Belle do that?

\vskip 0.5 truecm
\noindent
\underline{K. Seth}
I do not know what Belle's plans for the Super-$B$ Factory are, However, at
CLEO we are trying to access these states. But the cross sections are very
small. New projects like PANDA should certainly try.

\vskip 0.5 truecm
\noindent
\underline{E. Pallante}
Another comment on $g-2$, what might be feasible on the lattice is the
prediction of the hadronic component. From an experimental point view
is more important to reach an agreement between $e^+e^-$ and $\tau$ data.

\vskip 0.5 truecm
\noindent
\underline{S. Paul} 
This is a comment on Bugg's statement. I would like to comment that Compass
is still collecting data  and I would like to comment to Eberhardt 
that LEP has already thought
us how to combine data from different collaborations. And this requires
people from different collaborations to work together.

\vskip 0.5 truecm
\noindent
\underline{R. Faccini}
Theorists always want more data. The main challenge involves the $D$'s, there
we would need at least one hundred times more statistics.

\vskip 0.5 truecm
\noindent
\underline{R. Mussa}
Beauty factories are perfect tools for discovery, but detailed studies
ask for dedicated facilities. How high in energy will BES-III go?
Will it be possible to scan the $Y(4260,4350,4660)$ states recently
discovered and discussed at this conference? BES-III has good chances
to contribute to these studies. Concerning $p\overline{p}$ experiments like Panda,
so far we have no evidence of $X(3872)$ coupling to $p\overline{p}$, or to any other
baryon-antibaryon pair. Only the detection of the $X(3872)$ decay to any
baryon-antibaryon pair (for instance, $\eta_c$ preferentially decays in
lambda antilambda) would indicate that we have a realistic possibility
to study the newly discovered exotic states in a $p\overline{p}$ formation
experiment.

\vskip 0.5 truecm
\noindent
\underline{S. Glazek}
One main point of discussion is 
why quark model works, and if lattice can help with understanding that.
The major point is how
quarks and gluons lump into constituent quarks. 
In standard approaches one has nontrivial ground states to explain.
This question prompted Wilson and others
to develop a light-front approach to QCD.
I would say, lattice is not enough: once we have lattice data we need
some theory to interpret it.

\vskip 0.5 truecm
\noindent
\underline{L. Maiani}
Concerning constituent quarks, it seems to me that the audience is
divided in two parties, one has very clear idea about them, the other has
not.
So I suggest that they get together and discuss, and one part explain to
the other what to do.
Speaking for myself, I am in the middle: there are things we do
understand.
Concerning lattice, this is a field theoretic approach and
when we extract the masses from the propagator, this is a sound procedure
which does not need any further interpretation. Would Christine like to
comment on this?

\vskip 0.5 truecm
\noindent
\underline{C. Davies}
I do certainly agree.

\vskip 0.5 truecm
\noindent
\underline{S. Eidelman}
This is again on muon $g-2$, and the need of more data to resolve the discrepancy
between $\tau$ and $e^+e^-$.
There is no real problem, but still there is a puzzle to be resolved, but 
luckily we also see the outcome
of the data analysis from different groups, Kloe and Novosibirsk, 
from $e^+e^-$ coming together, 
and also new data from the $\tau$ decays in two pions from Belle
where we see that our results for the hadronic contributions to $g-2$ are 
in agreement with previous results from Cleo. Upcoming new machines, one already
being commissioned in Novosibirsk, continuing  a low energy $e^+e^-$ scan, up to 2 GeV, 
10 to 50 times better statistics, and of course I very strongly vote for DAFNE2. 
At the same time Belle and BaBar will help with $\tau$ decays. There is also a 
theoretical question, as to whether we understand well enough the SU(2) breaking 
corrections in $\tau$ decays.

\vskip 0.5 truecm
\noindent
\underline{F. J. Yndur\'ain}
You can fit $e^+e^-$ data and $\tau$ data together just allowing a slightly 
different $\rho$ mass and width. So I do not think there is any real disagreement 
to worry about.

\vskip 0.5 truecm
\noindent
\underline{H. Koch}
This is a comment on the
relevance of the proton antiproton physics for the search of 
new particles. It is true that so far we haven't seen a coupling 
of $p\overline{p}$ to these new states.  
In BaBar there might be a good chance to observe thee states.

\vskip 0.5 truecm
\noindent
\underline{K. Seth}
I would like to add one remark on the $p\overline{p}$ possibilities. Of course, a
new machine should try what it can., but the unfortunate fact is that the
coupling of these states to $p\overline{p}$ is going to be very small. We have
already seen that at Fermilab E760/E835. The pQCD prediction, a la
Brodsky, is that the coupling is  inversely proportional to the  eighth
power of the quark masses. So, the cross sections for populating heavy
quark states are going to be very much smaller than found for the light
quark states studied at LEAR.

\vskip 0.5 truecm
\noindent
\underline{L. Maiani}
One should keep in mind the difference between between the cross
section in $p\overline{p}$, the other is the inclusive production in
$p\overline{p}$ collisions. $p\overline{p}$ at Tevatron, we know the order
of magnitude of the cross section.
If we go to FAIR, there is a matter of energy. Any comment from FAIR?

\vskip 0.5 truecm
\noindent
\underline{U. Wiedner}
For Production of $J/\psi$ we  have about 120 nb.

\vskip 0.5 truecm
\noindent
\underline{L. Maiani}
As now it is time to conclude, I have basically  three points.

\begin{itemize}
\item We seem to be worried about 
applications of QCD to experiments. This is a very interesting 
message, with all the limitations QCD might have.
Feedback from theories to experiments for these kind of states
will be very interesting, but do not expect too much, since we do
not know how to solve QCD, we can make guesses, but sometimes guesses
might be wrong, although of course this does not kill the model.

\item Coming to the data, 
my personal worry is, which are the machines which will produce
the data? A Super$B$-factory might take some time. So I hope the issue
can be addressed at $p\overline{p}$ colliders, at the Tevatron, 
or maybe at FAIR.

\item
Finally, I think that a workshop on quark constituent masses will be 
a very appropriate outcome of this discussion.

\end{itemize}
\noindent
{\em This report was partly based on transcriptions of the recording,
necessarily shortened to remain within a reasonable page limit.
The Editors thank all the participants, apologize for any 
mistake and/or incomplete rendering of their contributions, and hope
that they nonetheless managed to convey the basic messages and the 
feeling of a very lively discussion.}

\end{document}